\documentclass[preprintnumbers,nofootinbib,noshowpacs,eqsecnum,twocolumn,prd]{revtex4}

\usepackage{graphicx}
\usepackage{amsmath,amsthm,amssymb}
\usepackage{mathrsfs,bbm}
\usepackage{color}

\definecolor{lgray}{gray}{0.9} 		
\graphicspath{ {figures/} }

\makeatletter       

\renewcommand{\p@subsection}{}

\makeatother

\newtheorem*{definition}{Definition}
\newtheorem*{algorithm}{Algorithm}

\newcommand*{\eweakgroup}{\mbox{$SU(2)_L \times U(1)_Y$} }
\newcommand*{\emgroup}{\mbox{$U(1)_{em}$} }

\newcommand*{\CP}{\mbox{\it CP} }

\newcommand*{\abs}[1]{\left\lvert {#1} \right\rvert} 
\newcommand*{\twomat}[1]{\underline{#1}}             
\newcommand*{\by}{\!\times\!}                        

\newcommand*{\dcp}{\delta_{\kappa}'}
\newcommand*{\dedm}{\delta_{\text{EDM}}}
\newcommand*{\tl}[0]{\succ}                          
\newcommand*{\tlex}[0]{\succ_{\text{lex}}}           
\newcommand*{\tdeg}[0]{\succ_{\text{deg}}}           
\newcommand*{\kx}[0]{K[\mathbf{x}]}
\newcommand*{\di}{\qquad}

\DeclareMathOperator{\sign}{sign}
\DeclareMathOperator{\normf}{normf}
\DeclareMathOperator{\lcm}{lcm}	
\DeclareMathOperator{\spol}{spol}		
\DeclareMathOperator{\lp}{LP}			
\DeclareMathOperator{\lc}{LC}			

\begin{document}

\preprint{HD-THEP-06-18}
\title{Determining the global minimum of Higgs potentials via Groebner bases\\
- applied to the NMSSM}

\author{M. Maniatis}
    \email[E-mail: ]{M.Maniatis@thphys.uni-heidelberg.de}
\author{A. von Manteuffel}
    \email[E-mail: ]{A.v.Manteuffel@thphys.uni-heidelberg.de}
\author{O. Nachtmann}
    \email[E-mail: ]{O.Nachtmann@thphys.uni-heidelberg.de}

\affiliation{
Institut f\"ur Theoretische Physik, Philosophenweg 16, 69120
Heidelberg, Germany
}


\begin{abstract}
Determining the global minimum of Higgs potentials with several
Higgs fields like the next-to-minimal supersymmetric extension of
the Standard Model (NMSSM) is a non-trivial task already at the tree level.
The global minimum of a Higgs potential
can be found from the set of all its stationary points
defined by
a multivariate polynomial system of equations.
We introduce here the algebraic Groebner basis approach to solve this system
of equations.
We apply the method to the NMSSM with \CP conserving as well as
\CP violating parameters.
The results reveal an interesting stationary-point structure of the potential.
Requiring the global minimum to give the electroweak symmetry breaking
observed in Nature excludes large parts of the parameter space.
\end{abstract}


\maketitle
\newpage

\newpage


\section{Introduction}
\label{intro}

It is a non-trivial task to find the global minimum for Higgs potentials
with a large number of Higgs fields. For instance in the 
next-to-minimal supersymmetric extension of the Standard Model
(NMSSM)~\cite{Fayet:1974pd}, the Higgs sector consists of
two (complex) electroweak doublets and one (complex) electroweak singlet,
that is 8 real fields from the doublets plus 2 real fields from the singlet.
The conventional approach based on the unitary gauge requires
the global minimum to be found in a 7-dimensional field space.

In this paper we introduce an algebraic approach
to determine the global minimum of the Higgs potential.
We describe how to compute all stationary points,
from which then the one with the lowest value of the potential
can be identified as the global minimum.
We apply the method to the NMSSM, where we can 
reveal a quite surprising structure of stationary points, that is
minima, maxima, and saddle points with different behaviour with respect to
the symmetry breaking of the \eweakgroup~electroweak gauge group.

The global minimum of the Higgs potential gives the expectation
values of the Higgs fields at the stable vacuum.
Parameter values for the Higgs potential are thus considered acceptable only,
if the global minimum of the Higgs potential occurs for Higgs
field vacuum expectation values, which induce the spontaneous breakdown
of \eweakgroup to the electromagnetic \emgroup at the observed electroweak
scale $v\approx 246$~GeV.

We consider the tree level Higgs potential for general models with two
Higgs doublets and an arbitrary number of additional Higgs singlets.
The first step is to notice that the potential is restricted by renormalisability
and gauge invariance. Renormalisability
requires at most quartic terms in the real fields in the potential. 
Electroweak gauge invariance restricts the possible doublet terms
in the potential, since
only gauge invariant scalar products of doublets can occur. 
Substituting the doublet fields by appropriate functions of
these invariant terms, we eliminate all gauge degrees of freedom
from the potential and effectively reduce the occurring powers in the doublet terms.
The method to base the analysis on quadratic gauge invariant functions
was introduced already in context with the general two-Higgs doublet model~\cite{Nagel:2004sw, Maniatis:2006fs}.

If the potential is bounded from below, the global minima are given by the stationary points with the lowest value of the potential.
The stationarity conditions form a non-linear, multivariate,
inhomogeneous polynomial system of equations of third order.
In this work we want to introduce a systematic approach to solve these
-- in general quite involved -- systems of polynomial equations.
Since we eliminate the gauge degrees of freedom by means of
gauge invariant functions the number of complex solutions
is assumed to be finite, which implies the absence of further continuous
symmetries.
We propose to determine the stationary points by a Groebner basis computation,
which is well established in ideal theory~\cite{Buchberger, Weispfenning, Bose}.
The Groebner basis was originally introduced to solve the {\em ideal membership problem}.
Constructing this Groebner basis in an appropriate order
of the {\em monomials} (the terms of the polynomials including coefficients),
for instance the {\em lexicographical ordering},
and subsequent triangularisation allows to solve the 
initial system of equations.
The method guarantees that all stationary points are found.

We apply the method introduced to the NMSSM.
For the computation of Groebner bases as well as the subsequent steps to solve
the systems of equations we employ the freely available open-source
algebra program \mbox{SINGULAR~\cite{Singular}}.
We find that large parts of the parameter space of the NMSSM
Higgs potential can be excluded
by requiring the global minimum to lead to the electroweak symmetry
breaking observed in Nature.
We illustrate this by determining the allowed and forbidden ranges
for some generic parameter values for this model.

In the literature we found many conditions
which constrain parts of the parameter space in the NMSSM, 
see for instance~\cite{Miller:2003ay, Ellis:1988er, Ellwanger:2004xm}.
Typically the conditions used are necessary but it is not
clear if they are also sufficient to ensure that the
resulting theory is acceptable.
We also want to mention that there is 
a purely numerical approach to
determine the global minimum in the 
effective one-loop
NMSSM Higgs potential~\cite{Funakubo:2004ka}.
The aim of our present work is to systematically 
reveal all stationary points of the tree level potential
by solving the system of equations which originates 
from the stationarity condition. 
We show, that this
can be done for the full parameter space including \CP violation.
With our method we can decide unequivocally if a given parameter
set of the tree level Higgs potential leads to an acceptable theory 
or not.

\section{The method}
\label{method}

We consider the tree-level Higgs potential of models
having \eweakgroup (weak isospin and hypercharge) electroweak
gauge symmetry. In particular we study models with
two Higgs doublets and $n$ additional real Higgs isospin and hypercharge
singlets.
This includes in particular THDMs, where we have no additional Higgs 
singlets, and the NMSSM with one additional complex Higgs singlet
corresponding to two real singlets.
We assume both doublets to carry hypercharge $y=+1/2$ and denote the
complex doublet fields by
\begin{equation}
\label{eq-doubldef}
\varphi_i(x) = \begin{pmatrix} \varphi^+_i(x) \\  \varphi^0_i(x) \end{pmatrix},
\quad
i=1,2 .
\end{equation}
For the singlets we assume real fields which we denote by
\begin{equation}
\phi_i(x), \quad i=1,\ldots,n .
\end{equation}

We remark that in supersymmetric models like the NMSSM the two Higgs doublets $H_d$, $H_u$
carry hypercharges \mbox{$y=-1/2$} and \mbox{$y=+1/2$} respectively.
This can be translated to the convention used here by setting
\begin{equation}
\label{eq-thdmsusytrafo}
\begin{split}
\varphi^\alpha_{1} & = - \epsilon_{\alpha \beta} ( H_d^{\beta} )^{*},\\
\varphi^\alpha_{2} & = H_u^\alpha,
\end{split}
\end{equation}
where
\begin{equation}
(\epsilon_{\alpha\beta}) = \begin{pmatrix} 0 & 1 \\ -1 & 0 \end{pmatrix}.
\end{equation}
Complex singlet fields are embedded in our notation by treating
the real and imaginary parts of the complex singlets as two real singlet
degrees of freedom.

\subsection{Gauge invariant functions}
\label{quadratics}

In the most general \eweakgroup gauge invariant Higgs potential
with the field content described above,
the doublet degrees of freedom enter solely via products of the
following form:
\begin{equation}
\label{eq-potterms}
\varphi_i^{\dagger}\varphi_j
\qquad \text{with } i,j \in \{1,2\}.
\end{equation}
It is convenient to discuss the properties of the potential such
as its stability and its stationary points in terms of
these gauge invariant quadratic expressions.
This was discussed in detail for THDMs and
also extended for the case of more than two doublets
in~\cite{Maniatis:2006fs}.
We recall the main steps here.

We arrange all possible \eweakgroup invariant
scalar products into the hermitian $2\by 2$~matrix
\begin{equation}
\label{eq-kmat}
\twomat{K} :=
\begin{pmatrix}
  \varphi_1^{\dagger}\varphi_1 & \varphi_2^{\dagger}\varphi_1 \\
  \varphi_1^{\dagger}\varphi_2 & \varphi_2^{\dagger}\varphi_2
\end{pmatrix}
\end{equation}
and consider its decomposition
\begin{equation}
\label{eq-kmatdecomp}
\twomat{K}_{i j} =
 \frac{1}{2}\,\left( K_0\,\delta_{i j} + K_a\,\sigma^a_{i j}\right),
\end{equation}
where $\sigma^a$ are the Pauli matrices.
The four real coefficients in this decomposition are
\begin{equation}
\label{eq-kdef}
K_0 = \varphi_{i}^{\dagger} \varphi_{i},
\quad
K_a = ( \varphi_{i}^{\dagger} \varphi_{j} )\, \sigma^a_{ij} ,
\quad a=1,2,3,
\end{equation}
where here and in the following summation over repeated indices is understood.
The matrix~(\ref{eq-kmat}) is positive semi-definite, which implies
\begin{equation}
\label{eq-kconditions}
K_0 \ge 0, \quad K_0^2-K_1^2-K_2^2-K_3^2 \ge 0.
\end{equation}
On the other hand, for every hermitian $2\by 2$~matrix $\twomat{K}_{i j}$
satisfying~\eqref{eq-kconditions} there exist fields~\eqref{eq-doubldef}
satisfying~\eqref{eq-kmat}, see~\cite{Maniatis:2006fs}.
It was also shown in~\cite{Maniatis:2006fs} that the four quantities
$K_0,K_a$ satisfying~\eqref{eq-kconditions} parametrise the
gauge orbits of the Higgs doublets.
Using the inversion of~(\ref{eq-kdef}),
\begin{equation}
\label{eq-phik}
\begin{alignedat}{2}
\varphi_1^{\dagger}\varphi_1 &= (K_0 + K_3)/2, &\quad
\varphi_1^{\dagger}\varphi_2 &= (K_1 + i K_2)/2, \\ 
\varphi_2^{\dagger}\varphi_2 &= (K_0 - K_3)/2, &
\varphi_2^{\dagger}\varphi_1 &= (K_1 - i K_2)/2,
\end{alignedat}
\end{equation}
we can replace the doublet terms of the potential 
-- due to renormalisability --
by at most quadratic terms in the real functions $K_0$, $K_1$,
$K_2$, and $K_3$, which simplifies the potential and
eliminates all \eweakgroup~gauge degrees of freedom.
We thus end up with a potential of the form
\mbox{$V(K_0, K_1, K_2, K_3, \phi_1,\ldots,\phi_n)$}.

\subsection{Stationary points}
\label{stationarity}
To determine the stationary points of the Higgs potential
we consider \mbox{$V(K_0, K_1, K_2, K_3, \phi_1,\ldots,\phi_n)$}
and take the constraint \eqref{eq-kconditions} into account.
We distinguish the possible cases of stationary points
by the \eweakgroup symmetry breaking behaviour
which a vacuum of this type would have (see~\cite{Maniatis:2006fs}):
\begin{itemize}\label{eq-kcondnon}
\item{{\bf unbroken~\eweakgroup}: A stationary point with
\begin{equation}
K_0=K_1=K_2=K_3=0.
\end{equation}
A global minimum of this type implies vanishing vacuum expectation
value for the doublet fields~(\ref{eq-doubldef})
and therefore the trivial behaviour with respect to the gauge group.
The stationary points of this type are found by setting 
all Higgs-doublet fields (or correspondingly the $K_0$ as well as
the $K_a$ fields) in the potential to zero and requiring a vanishing
gradient with respect to the remaining real fields:
\begin{equation}\label{eq-stationarityU}
\nabla \; V(\phi_1, \ldots, \phi_n) =0.
\end{equation}
}
\item{{\bf fully broken~\eweakgroup}: A stationary point with
\begin{equation}\label{eq-kcondfull}
\begin{gathered}
K_0>0,\\
K_0^2-K_1^2-K_2^2-K_3^2 > 0.
\end{gathered}
\end{equation}
A global minimum of this type has non-vanishing
vacuum expectation values for the
charged components of the doublets fields in~(\ref{eq-doubldef}),
thus leads to a fully broken~\eweakgroup, see~\cite{Maniatis:2006fs}.
The stationary points of this type are found by requiring a vanishing gradient
with respect to all singlet fields and all gauge invariant functions:
\begin{equation}
\label{eq-stationarityF}
\nabla \; V(K_0, K_1, K_2, K_3, \phi_1, \ldots, \phi_n) =0.
\end{equation}
The constraints~\eqref{eq-kcondfull} on the gauge invariant functions
must be checked explicitly for the real solutions found.
}
\item{{\bf partially broken~\eweakgroup}: A stationary point with
\begin{equation}\label{eq-kcondpartial}
\begin{gathered}
K_0>0,\\
K_0^2-K_1^2-K_2^2-K_3^2 = 0.
\end{gathered}
\end{equation}
For a global minimum of this type follows the desired partial breaking
of~\eweakgroup down to~\emgroup, see~\cite{Maniatis:2006fs}.
Using the Lagrange method, these stationary points are given
by the real solutions of the system of equations
\begin{equation}\label{eq-stationarityP}
\begin{split}
\nabla
\big[
V(K_0, K_1, K_2, K_3, \phi_1, \ldots, \phi_n)\quad& \\
       - u \cdot (K_0^2-K_1^2-K_2^2-K_3^2)\; \big] &= 0,
\\
K_0^2-K_1^2-K_2^2-K_3^2 &= 0,
\end{split}
\end{equation}
where $u$ is a Lagrange multiplier.
The inequality in~\eqref{eq-kcondpartial} must be
checked explicitly for the solutions found for~\eqref{eq-stationarityP}.
}
\end{itemize}
For a potential which is bounded from below,
the global minima will be among these stationary points.
Solving the systems of equations~\eqref{eq-stationarityU},
\eqref{eq-stationarityF}, and \eqref{eq-stationarityP}, and inserting
the solutions in the potential, we can therefore
identify the global minima as those solutions which have the lowest value of
the potential.
Note that in general there can be more than one global minimum point.

From the mathematical point of view we have
with~(\ref{eq-stationarityU},
\ref{eq-stationarityF}, \ref{eq-stationarityP})
to solve non-linear, multivariate, inhomogeneous systems 
of polynomial equations 
of third order.
We want to demonstrate that this is possible, even if the number of
fields is large, like in the NMSSM. The most involved case is given
by~(\ref{eq-stationarityP}), which for the NMSSM consists of seven equations
in seven indeterminates, namely six real fields and one Lagrange
multiplier.
We stress that at this point we assume the number of complex
solutions of ~(\ref{eq-stationarityU}, \ref{eq-stationarityF},
\ref{eq-stationarityP}) to be finite, that is, we have
systems of equations with a finite number of solutions.
The gauge invariant functions avoid spurious continuous sets of
complex solutions, which we found to arise in the case of the
MSSM as well as the NMSSM if the stability conditions are
formulated with respect to the Higgs fields~\eqref{eq-doubldef}
in a unitary gauge.

The solution of multivariate polynomial systems of equations
is the subject of polynomial ideal theory and can
be obtained algorithmically in the Groebner basis approach~\cite{Buchberger}. 
See appendix~\ref{ap-buchberger} for a brief introduction to
this subject.
Within this approach the system of equations is transformed into
a unique standard form with respect to
a specified underlying ordering of the polynomial summands ({\em monomials}).
This unique standard form of the system of equations is given by the
corresponding {\em reduced Groebner basis}.
If the underlying order is the {\em lexicographical} ordering,
the unique standard form consists of equations with a partial separation
in the indeterminates.
We use a variant of the $F_4$ algorithm~\cite{Faugere-F4} to compute the Groebner bases.
A Groebner basis computation is generally much faster if the standard form is computed with respect to {\em total degree} orderings and 
then transformed into a lexicographical Groebner basis.
The transformation of bases from total degree to lexicographical ordering
is done with the help of the FGLM algorithm~\cite{FGLM}.
Finally, the system of equations represented by the lexicographical Groebner basis
has to be triangularised.
The decomposition of the system of equations with a
finite number of solutions into triangular sets is 
performed with the algorithm introduced in~\cite{Moeller}.
Each triangular system consists of one univariate equation, 
one equation in $2$ indeterminates, one equation in $3$ indeterminates
and so forth. This means that the solutions are found
by subsequently solving just univariate equations by inserting the
solutions from the previous steps.

The construction of the Groebner basis as well as the 
triangularisation are done algebraically, 
so no approximations are needed. 
However, the triangular system of equations 
contains in general polynomials of high order, where the zeros
cannot be obtained algebraically.
Here numerical methods are needed to find the in general complex roots
of the univariate polynomials. 

In more involved potentials, like the NMSSM, the algorithmic solution
is considerably simplified (or even made accessible), if the coefficients
of the polynomials are given in form of rational numbers.
Since rational numbers are
arbitrarily close to real numbers and moreover the physical parameters are
given only with a certain precision this does not limit the
general applicability of the method in practice.

All algorithms for the computation of the Groebner basis with respect to a given order
of the monomials, the change of the underlying order, the 
triangularisation, and the solution of the triangular systems 
are implemented in the SINGULAR program package~\cite{Singular}.
The solutions obtained can be easily checked by inserting them into the 
initial system of equations. Moreover, the number of complex solutions, that is the
multiplicity of the system, is known,
so we can easily check that no stationary point is missing.

\section{Stationary points in the NMSSM}
\label{NMSSM}

Now, we want to apply the methods introduced in
section~\ref{method} to the NMSSM.

\subsection{The NMSSM Higgs potential}

The NMSSM Higgs sector contains two doublets and one singlet,
\begin{equation}
H_u=
  \begin{pmatrix}
  H_u^+\\
  H_u^0
  \end{pmatrix},
\qquad
H_d=
  \begin{pmatrix}
  H_d^0\\
  H_d^-
  \end{pmatrix},
\qquad
S,
\end{equation}
with the tree level Higgs
potential
\mbox{$V_{\text{NMSSM}}=V_{\text{F}}+V_{\text{D}}+V_{\text{soft}}$}~\cite{Ellis:1988er, Miller:2003ay}, where
\begin{align}
\begin{split}
\label{eq-NMSSMpot}
V_{\text{F}} = & 
|\lambda S|^2 (|H_u|^2+|H_d|^2) + |\lambda H_uH_d +\kappa S^2|^2, \\
V_{\text{D}} = &
\frac{1}{8} \bar g^2 (|H_d|^2-|H_u|^2)^2
 +\frac{1}{2}g^2|H_u^{\dagger}H_d|^2, \\
V_{\text{soft}} = & 
m_{H_u}^2|H_u|^2 + m_{H_d}^2|H_d|^2 + m_S^2|S|^2\\
&+ [\lambda A_{\lambda}SH_uH_d+\frac{1}{3}\kappa A_{\kappa} S^3+\text{h.c.} ].
\end{split}
\end{align}
We use the notation
$H_u H_d \equiv \epsilon_{\alpha \beta}(H_u)^{\alpha}(H_d)^{\beta}=H_u^+H_d^--H_u^0H_d^0$
and $\bar g = \sqrt{g^2+g^{\prime 2}}$, where 
$g$ and $g^{\prime}$ are the $SU(2)_L$ and $U(1)_Y$ gauge couplings, respectively.
The parameters of the potential are given by the experimentally 
fixed electroweak gauge couplings
and
\begin{equation}
\label{eq-par}
\lambda, \kappa, m_{H_u}^2, m_{H_d}^2, m_{S}^2, A_{\lambda}, A_{\kappa}.
\end{equation}

The quartic terms of the potential~(\ref{eq-NMSSMpot})
are positive for any non-trivial field configuration,
if both $\lambda, \kappa$ are non-vanishing.
The potential is therefore bounded from below for all cases considered here,
and stability has not to be checked any further.

We translate the NMSSM Higgs potential to the formalism described
in the previous section, where we decompose the complex singlet field into
two real fields according to \mbox{$S=S_{re} + i S_{im}$}.
In this notation the potential is given by
\begin{align}
\begin{split}
\label{eq-NMSSMpot2}
V_{\text{F}} =&\;
  \frac{1}{4} |\lambda|^2 \left( K^2_1+K^2_2 + 4 K_0 (S^2_{re}+ S^2_{im}) \right)\\
&+ |\kappa|^2 ( S^2_{re}+S^2_{im} )^2\\
&- \operatorname{Re}(\lambda \kappa^*) 
	\left(K_1 (S_{re}^2-S_{im}^2) + 2 K_2 S_{re}S_{im}\right)\\
&+ \operatorname{Im}(\lambda \kappa^*) 
	\left(K_2 (S^2_{re}-S^2_{im} )- 2 K_1 S_{re}S_{im}\right),\\
V_{\text{D}} =&\;
   \frac{1}{8}\bar{g}^2 K_3^2 
+ \frac{1}{8} g^2 \; \left( K^2_0-K^2_1-K^2_2-K^2_3 \right),\\
V_{\text{soft}} =&\;
  \frac{1}{2} m_{H_u}^2 \; (K_0-K_3) 
+ \frac{1}{2} m_{H_d}^2 \; (K_0+K_3)\\ 
&+ m_{S}^2 \; (S^2_{re} + S^2_{im})\\ 
&- \operatorname{Re}(\lambda A_{\lambda})
	 \left( K_1 S_{re} - K_2 S_{im} \right)\\
&+ \operatorname{Im}(\lambda A_{\lambda})
	 \left( K_2 S_{re} + K_1 S_{im} \right)\\
&+ \frac{2}{3} \operatorname{Re}(\kappa A_{\kappa})
	 \left( S_{re}^3 - 3 S_{re} S_{im}^2 \right)\\
&+ \frac{2}{3} \operatorname{Im}(\kappa A_{\kappa})
	 \left( S_{im}^3 - 3 S_{re}^2 S_{im} \right).
\end{split}
\end{align}
For given values of the potential parameters \eqref{eq-par}
we can find all stationary points of the NMSSM by 
solving the systems 
of equations~(\ref{eq-stationarityU}, \ref{eq-stationarityF}, \ref{eq-stationarityP})
as described above.

\subsection{Choice of parameters}

In order to fix experimentally known parameters like the electroweak scale
it is inappropriate to choose numerical values for the parameter set~\eqref{eq-par}.
Instead, we express different original parameters in terms of the desired vacuum
expectation values of the neutral components of the Higgs doublets and the Higgs singlet,
the mass of the charged Higgs boson and a \CP-violating phase.
To this end we introduce the vacuum expectation values
of the Higgs fields,
\begin{align}
\begin{split}
 &\left<H_u \right> =
 e^{i\varphi_u}  \begin{pmatrix}
  0\\ \frac{1}{\sqrt{2}} v_u
 \end{pmatrix},\\
 &\left<H_d \right> =
 \begin{pmatrix}
  \frac{1}{\sqrt{2}} v_d\\ 0
 \end{pmatrix},\\
 &\left<S \right> = \frac{1}{\sqrt{2}}e^{i\varphi_S} v_S,
\end{split}
\end{align}
which are parametrised by $v_u$, $v_d$, $v_S$, and the phases
$\varphi_u$, and $\varphi_S$.
The stationarity condition of the potential at the vacuum 
values for the fields
requires a vanishing gradient with respect to the Higgs fields.
These equations relate the soft-breaking mass parameters
with the vacuum expectation values.
As usual, we define $v^2 \equiv v_u^2+v_d^2$ and $\tan \beta \equiv v_u/v_d$.
Further, if we write the complex parameters
$\lambda$, $\kappa$, $A_{\lambda}$, and $A_{\kappa}$ in polar coordinates
with phases $\delta_\lambda$, $\delta_\kappa$, $\delta_{A_\lambda}$, $\delta_{A_\kappa}$
and introduce the abbreviations
\begin{equation}
\dedm \equiv \delta_\lambda+\varphi_u+\varphi_S, \quad
\dcp \equiv \delta_\kappa+3\varphi_S,
\end{equation} 
the initial parameters of the potential~(\ref{eq-par}) can be replaced by the
new set of parameters
\begin{equation}
\label{eq-parn}
\lambda, \kappa, |A_\kappa|, \tan \beta, 
v_S, m_{H^\pm}, \sign{ R_\kappa }, \dedm, \dcp. 
\end{equation}
plus the electroweak scale $v\approx 246$~GeV.
Note, that it is not sufficient to supply the length $|A_\kappa|$, 
in addition we have to fix the sign of
\begin{equation}
R_\kappa \equiv \frac{1}{\sqrt{2}} \operatorname{Re}
 \left( \kappa A_\kappa e^{i 3\varphi_S}\right).
\end{equation}
In the mass matrix of the Higgs scalars, the \CP violating entries which mix the ``scalar''
with the ``pseudoscalar'' fields are proportional to
the imaginary part of $\exp [i (\dedm - \dcp) ]$.


\subsection{Numerical results}

As a numerical example we choose the parameter values
\begin{equation}\label{eq-centralvals}
\begin{gathered}
  \lambda = 0.4, \quad
  \kappa = 0.3, \quad
  \abs{A_\kappa}  = 200\text{~GeV},\\
  \tan\beta = 3,\quad
  v_S = 3 v, \quad
  m_{H^\pm} = 2 v,\\
  \sign{R_\kappa}=-, \quad
  \dedm = 0, \quad
  \dcp = 0
\end{gathered}
\end{equation}
and consider the variation of one parameter at a time with
the values of the other parameters in~\eqref{eq-centralvals} kept fixed.
For a given point in parameter space we compute
all stationary points of the NMSSM potential as described
above.
As mentioned in section~\ref{stationarity} the Groebner basis construction
is performed with numerical coefficients. Here we use a 
precision of $12$ digits for the input parameters~(\ref{eq-par}), which
are determined from the values for the parameters~\eqref{eq-parn}.
The roots of the univariate polynomials are found numerically,
where we choose a precision of $100$ digits.
Our approach allows to use arbitrary precisions in both cases.
We verify that the errors of the approximate statements
described in the following are under control.
For generic values of the parameters we find $52$ complex solutions:
$7$ corresponding to the unbroken, $38$ to the partially broken, 
and $7$ to the fully broken cases.
The number of real and therefore relevant solutions depends on the
specific values of the parameters.

As expected from the $Z_3$ symmetry of the potential, we find either 1 or 3
solutions sharing the same value of the potential within
the accuracy of the numerical roots.
From the computed stationary points only those may be
accepted as global minima 
which correspond to the initial vacuum expectation values 
\pagebreak
(up to the complex phases), that is which fulfil
\begin{equation}
\label{eq-vac}
\sqrt{2 K_0} \approx v,\;\;
\sqrt{\frac{K_0-K_3}{K_0+K_3}} \approx \tan\beta,\;\;
\sqrt{2 ({S_{re}^2+S_{im}^2})} \approx v_S.
\end{equation}
Since for non-vanishing $\lambda, \kappa$ the potential is bounded from below,
the stationary point with the lowest value of the potential is
the global minimum.
Further, we determine for every stationary solution, whether it is
a local maximum, local minimum or a saddle point.
For the regular solutions, i.e. the partially- and fully-broken cases, this
is achieved via the bordered Hessian 
matrix~(see for instance~\cite{Fletcher})
in terms of 
$K_0, K_1, K_2, K_3, S_{re},S_{im}$.
This takes all powers of the doublet fields into account, which allows for a definite
decision on the type of the stationary point
also for frequently encountered partial breaking solutions
where at least one mass squared is zero and the others have the same sign.
For irregular solutions, i.e. the non-breaking solutions with $K_0=0$,
the Lagrange formalism can not be used since the gradients of the two boundary conditions
with respect to $K_0, K_1, K_2, K_3,S_{re},S_{im}$ become linearly dependent. Instead we resubstitute
the original fields $H_u^+,H_u^0,H_d^0,H_d^-,S_{re},S_{im}$ in this case and
consider the free Hessian matrix with respect to these fields. 
This turns out to be
sufficient in practice to judge on the type of the stationary points.

\begin{figure*}
\begin{center}
\includegraphics[height=0.633\textheight,width=\linewidth]{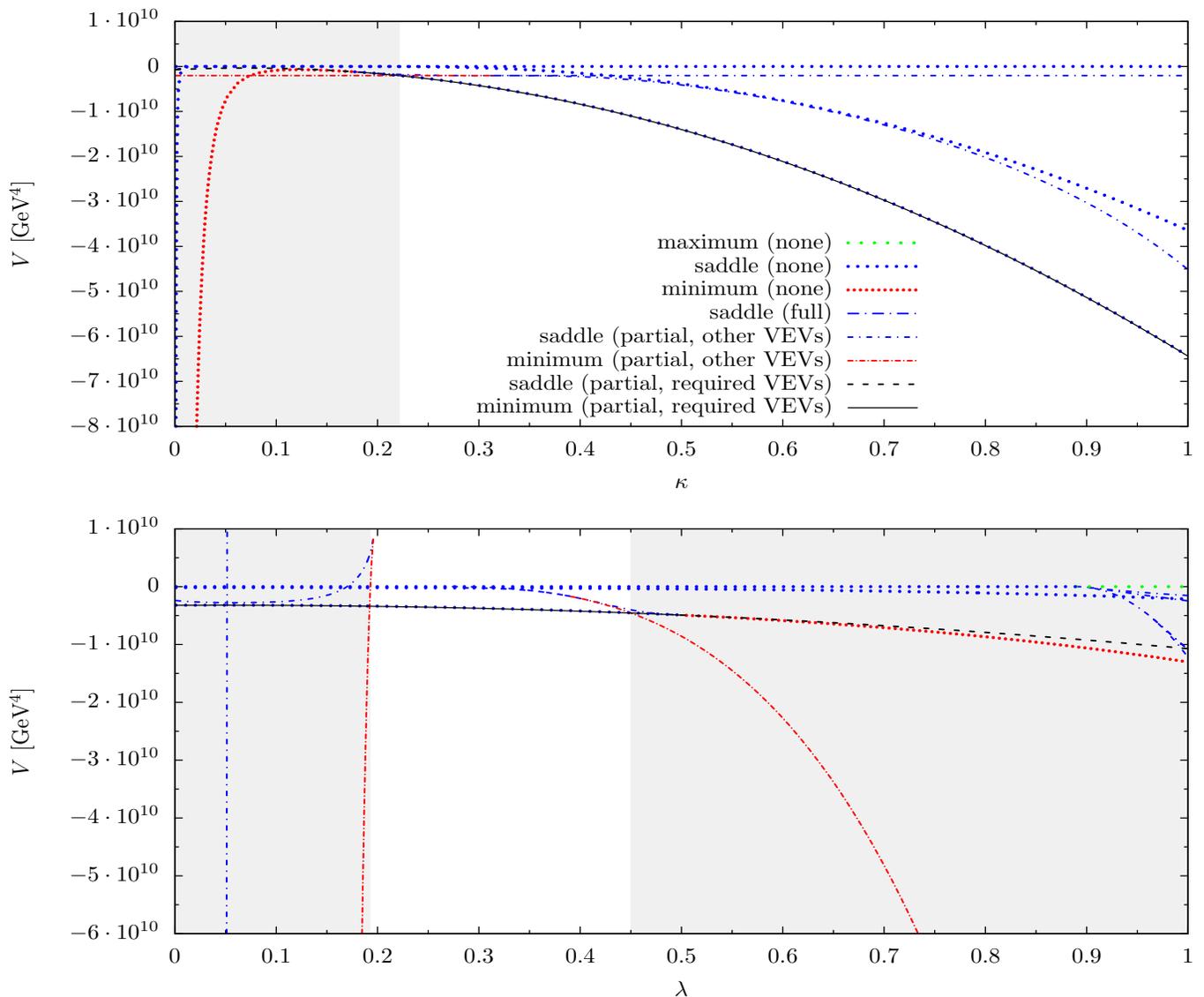}
\end{center}
\caption{\label{fig-statpnts2a}\label{fig-lambda}\label{fig-kappa}
Values of the NMSSM potential at its stationary points in dependence
of $\kappa,\lambda$. The following parameters are kept constant
unless explicitly varied:
$\lambda=0.4, \kappa=0.3, \abs{A_\kappa}=200\text{ GeV},
\tan\beta=3, v_S = 3 v, m_{H^\pm} = 2 v, \sign{R_\kappa} = -,
\dedm=\dcp=0$.
Each line corresponds to 1 or 3 stationary points sharing the same
value of the potential.
The different line styles denote saddle points, maxima, and minima.
The labels
'none', 'full', and 'partial' denote 
solutions of the classes with
unbroken~(\ref{eq-stationarityU}), 
fully broken~(\ref{eq-stationarityF}), and
partially broken~(\ref{eq-stationarityP})
\eweakgroup, respectively.
For solutions of the partially broken class, it is also denoted
whether they correspond to the 'required VEVs' $v_u,v_d,v_S$ or
to 'other VEVs'.
Excluded parameter values, where the global minimum does not exhibit
the required vacuum expectation values, are drawn shaded.
}
\end{figure*}

\begin{figure*}
\begin{center}
\includegraphics[totalheight=0.945\textheight,width=\linewidth]{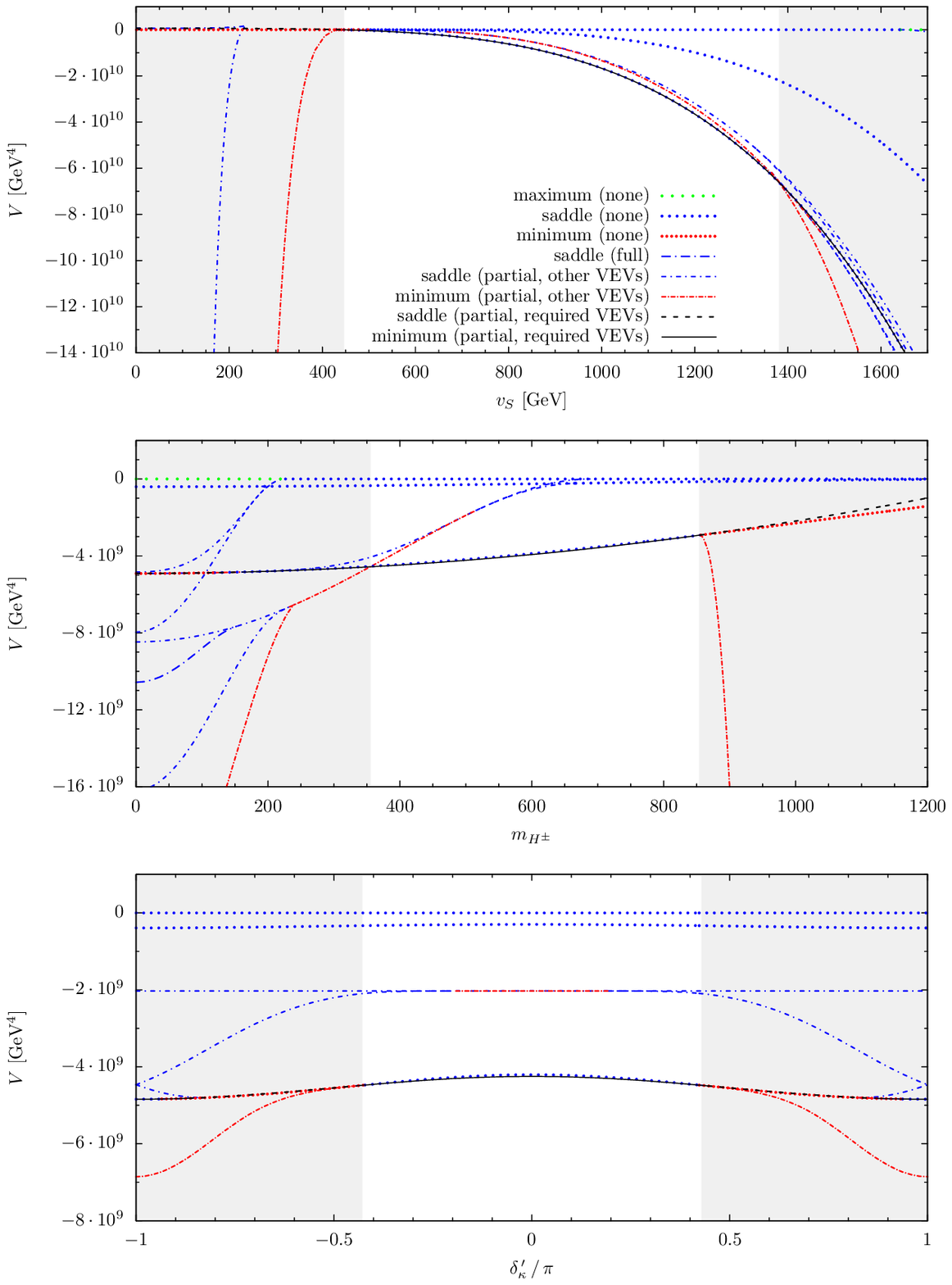}
\end{center}
\caption{\label{fig-statpnts2b}\label{fig-vs}\label{fig-mcha}\label{fig-delta}
Same as in Fig.~\ref{fig-statpnts2a} but for variation 
of $v_S, M_{H^\pm},\dcp$, respectively.
}
\end{figure*}

Figures~\ref{fig-lambda} and \ref{fig-delta} show the values of the potential
at all stationary points for the parameter values~(\ref{eq-centralvals}) 
and the cases where successively one of the parameters
$\lambda, \kappa, v_S, m_{H^\pm}, \dcp$ is varied.
Each curve in the Figures represents 1- or 3-fold degenerate stationary potential
values, where the gauge symmetry breaking behaviour of the solutions is denoted by
different line styles.
Excluded parameter regions, where the global minimum does not exhibit the
required expectation values~(\ref{eq-vac}) are drawn shaded.
As is illustrated by the figures we find that substantial regions
of the NMSSM parameter space are excluded.
For some excluded parameter regions, the partially breaking solutions
with the required vacuum expectation values~(\ref{eq-vac})
are saddle points (see for instance Fig.~\ref{fig-lambda}, top).
This means they can be discarded as global minima without
calculation of the other stationary points.
However, this is not always the case.
Obviously from Fig.~\ref{fig-vs}, top, we find an upper bound
for $v_S$.
For the plotted $v_S$ larger than this
upper exclusion bound the solutions fulfilling~(\ref{eq-vac})
are still pronounced minima, i.e. the mass
matrices have positive eigenvalues, but they are no
longer the global minima.
The influence of the \CP violating phase $\dcp$ is
shown in Fig.~\ref{fig-delta}, bottom.
Note that $\dcp \rightarrow -\dcp$ is not 
a symmetry of the potential.
However, the potential is invariant under
$(\dcp,K_2,S_{im})\rightarrow -(\dcp,K_2,S_{im})$,
that is $(\dcp,H_u,H_d,S)\rightarrow (-\dcp,H_u^\ast,H_d^\ast,S^\ast)$,
if the residual phases are chosen as in~(\ref{eq-centralvals}).
Therefore the stationary values of the potential in Fig.~\ref{fig-delta}
depend only on $\abs{\dcp}$.
Also we want to note that in all figures shown here there is
a non-breaking stationary point with potential values slightly above that of the 
{\em wanted} global minimum. 
We find that this effect is not coincidental for the
parameters~(\ref{eq-centralvals}) chosen here, but rather a generic feature of the NMSSM.
Within the \CP conserving parameter range
\begin{equation}\label{eq-randomrange}
\begin{gathered}
\lambda \in \,]0,1],\quad \kappa \in \,]0,1],\quad A_\kappa \in \pm\,]0,2500]\text{ GeV},\\
\tan\beta \in \,]0,50],\quad  v_s \in \,]0,5000]\text{ GeV},\\
m_{H^\pm} \in \,]0,2500]\text{ GeV}
\end{gathered}
\end{equation}
we find a typical separation of the potential values for these non-breaking solutions
and the solutions with the required vacuum expectation values~(\ref{eq-vac})
below the percent level, in many cases much smaller.
We do not find fully breaking global minima for scenarios in the range~\eqref{eq-randomrange}
where the solutions with the required vacuum expectation values~(\ref{eq-vac}) are local minima.
Eventually, we find examples, 
where \CP conserving parameter values with the ``wrong'' global minimum
produce the wanted global minimum 
if a non-vanishing phase $\dcp$ is introduced.

\onecolumngrid
\vfill
\pagebreak
\twocolumngrid
\section{Conclusion}
The stationarity conditions of Higgs potentials at the tree-level lead
in general to non-linear, multivariate, polynomial systems of equations.
We have formulated the stationarity conditions by means of gauge invariant
functions, thus avoiding to have to deal with electroweak gauge degrees of freedom.
We have introduced the Groebner basis approach to solve the 
systems of equations and have applied the method to the NMSSM.
Within the Groebner basis approach we have easily found all
stationary solutions even with general \CP violating parameters.
We have demonstrated that large regions of the parameter space of the NMSSM are
excluded because they do not lead to the required global minimum.
Finally, we note that the method proposed here should also be applicable to
other potentials which neither have to be restricted to
two doublets nor even have to be renormalisable.


\begin{acknowledgments}
We thank R.~Barbieri and  W.~Wetzel
for valuable discussions.
This work was supported by the German Bundesministerium f\"ur Bildung
und Forschung, project numbers HD05HT1VHA/0 and HD05HT4VHA/0.
\end{acknowledgments}


\appendix
\section{Buchberger algorithm}
\label{ap-buchberger}

In this appendix we want to sketch the construction of the Buchberger algorithm which
transforms a given set of polynomials~$F$ into a Groebner basis~$G$. 
The Groebner basis~$G$ has exactly the same simultaneous zeros
as the initial set of polynomials~$F$, but allows better access to the actual calculation
of these zeros.
The general idea is to {\em complete} the set~$F$ by adjoining differences of polynomials.
Before we can present the algorithms themselves we have to 
introduce the two basic ingredients, that is
{\em Reduction} and the
{\em S-polynomial}. 
For a more detailed discussion we refer the reader to
the literature~\cite{Buchberger,Weispfenning, Bose}. 
Here we follow closely~\cite{Bose}.
First of all we recall some definitions.
\begin{definition}{Polynomial Ring\\}
	A Polynomial Ring $K[x_1,\ldots,x_n]\equiv K[\mathbf{x}]$
 	is the set of all $n$-variate polynomials with variables $x_1,\ldots,x_n$
	and coefficients in the field $K$.
\end{definition}
\begin{definition}{Generated Ideal\\}
	Let $F=\{f_1,\ldots,f_n\} \subset K[\mathbf{x}]$ be finite,
	$F$ generates an ideal defined by\\
$
	I(F)\equiv\bigg\{ \sum\limits_{f_i \in F} r_i \cdot f_i
  \;\bigg|\;
  r_i \in K[\mathbf{x}],\;
  f_i \in F,\;
  i=1,\ldots,n\bigg\}.
$
\end{definition}
In the following we want to consider an explicit example, 
that is a set 
$F=\{f_1,f_2,f_3\} \subset \mathbbm{Q}[x,y]$
of polynomials with rational coefficients:
\begin{align}
\label{eq-poly}
	\begin{split}
	f_1 &= 3 x^2y+2xy+y+9x^2+5x-3,\\
	f_2 &= 2 x^3y-xy-y+6x^3-2x^2-3x+3,\\
	f_3 &= x^3y+x^2y+3x^3+2x^2.
	\end{split}
\end{align}
The set $F$ generates an ideal~$I(F)$, which is given
by the set of sums of $f_1$, $f_2$, and $f_3$, where each polynomial is 
multiplied with another arbitrary polynomial from the ring $\mathbbm{Q}[x,y]$.
The summands of the polynomial are denoted as {\em monomials}
and each monomial is the 
product of a coefficient and a {\em power product}.

Further, we introduce an {\em ordering} ($\tl$) of the monomials.
In the {\em lexicographical ordering}~($\tlex$) the monomials are ordered with respect
to the power of each variable subsequently.
The ring notation~$\mathbbm{Q}[x,y]$
defines $y \tlex x$, that is for the
lexicographical ordering of monomials powers of $y$ are considered first,
then powers of $x$.
Explicitly, this means $2 x^2 y^3 \tlex 5 x y^2  $ because
the power of $y$ is larger in the first monomial and 
$2 x y^2 \tlex 5 y^2$, because both monomials have the same power of~$y$,
but the first monomial has a larger power of~$x$. 
The monomials of the polynomials~(\ref{eq-poly})
from the ring $\mathbbm{Q}[x,y]$ 
are ordered with respect to lexicographical ordering.
In {\em total degree ordering}~($\tdeg$) the monomials are ordered with respect
to the sum of powers in each monomial. 
If two monomials have the same sum of powers,
they are ordered with respect to another ordering, for instance lexicographical.
For polynomials in $\mathbbm{Q}[x,y]$ we have
$x^2 y \tdeg 4 x y$ since the sum of powers of the left power product is $3$ compared to
$2$ for the right power product. 

The largest power product with respect to the underlying  
ordering ($\tl$) of a polynomial~$f$ is denoted as the
{\em leading power product}, $\lp(f)$, the corresponding coefficient as
{\em leading coefficient}, $\lc(f)$. 
With help of these preparations
we can define the two essential parts of the Buchberger algorithm, 
that is {\em Reduction} and the {\em S-polynomial}.
\begin{definition}{Reduction}\\
Let $f, p \in \kx$. We call $f$ reducible modulo $p$,
if for a power product $t$ of $f$ there exists a power product $u$ with 
$\lp(p) \cdot u=t$. 
Then we say, $f$ reduces to $h$
modulo $p$, where
$h= f - \frac{\text{Coefficient}(f,t)}{\lc(p)} \cdot u \cdot p$.
\\
\end{definition}
In the example~(\ref{eq-poly}) 
the polynomial~$f_3$ is reducible modulo $f_1$, 
since for example the second monomial of~$f_3$, that is $x^2 y$, is
a multiple of the $\lp(f_1)$, and $h=f_3 - 1/3 f_1 = x^3y -2/3 x y-1/3y+3x^3-x^2-5/3x+1$.

Reduction of a polynomial modulo a set $P \subset \kx$
is accordingly defined if there is a $p \in P$ such that
$f$ is reducible modulo $p$. Further, we say, a polynomial~$h$
is in {\em reduced form} or {\em normal form}
modulo~$F$, in short $\normf(h,F)$,
if there is no $h'$ such that $h$ reduces to $h'$ modulo~$F$.
A  set $P \subset \kx$ is called reduced,
if each $p\in P$ is in reduced form modulo $P \backslash \{ p \}$.
Note that reduction is defined with respect to
the underlying ordering of the monomials, since the
leading power product is defined with respect to the ordering.
In general, a normal form is not unique, neither for a polynomial
nor for a set.

Now we can present an algorithm, to compute 
a normal form $Q \subset \kx$ of a finite $F \subset \kx$. 
\begin{algorithm}{Normal form}\\
For a given finite set~\mbox{$F \subset \kx$} determine a
normal form~\mbox{$Q \subset \kx$}.\\[0pt]
\begin{center}
\colorbox{lgray}{\parbox{0.4\textwidth}
	{
\begin{align*}
	&Q:=F\\
& \text{{\bf while} exists } p \in Q\\
& \text{which is reducible modulo } Q \backslash \{p \} \;\;\text{\bf do}\\
&	\di Q:=Q \backslash \{ p \}\\
&	\di h:= \normf(p, Q)\\
&	\di \text{\bf if } h  \neq 0 \;\;\text{\bf then}\\
&	\di \di Q:=Q \cup \{ h \}\\
&	\text{\bf return } Q
\end{align*}
	}}
\end{center}
\end{algorithm}

\begin{definition}{S-polynomial}\\
For $g_1, g_2 \in \kx$ the S-polynomial of $g_1$ and $g_2$ is defined as
\begin{align*}
\spol(g_1,g_2) &\equiv
\frac{\lcm \big( \lp(g_1), \lp(g_2) \big)}{\lp(g_1)} g_1\\
&\quad-\frac{\lc(g_1)}{\lc(g_2)} \frac{\lcm \big( \lp(g_1), \lp(g_2) \big)}{\lp(g_2)} g_2,
\end{align*}
where $\lcm$ denotes the least common multiple.
\end{definition}

In the example~(\ref{eq-poly}) 
we can build the S-polynomial for any two polynomials, for instance
\begin{align*}
\spol(f_1,f_2) &= 
\frac{x^3y}{x^2y}\; f_1 - \frac{3}{2}\; \frac{x^3y}{x^3y}\; f_2
	= x\; f_1 - 3/2 \;f_2\\
&=2x^2y+5/2xy+3/2y+8x^2+3/2x-9/2.
\end{align*}

Finally we define the Groebner basis.
\begin{definition}{Groebner basis}\\
$G \subset \kx$ is called Groebner Basis,
if for all $f_1, f_2 \in G$ $\normf(\spol(f_1, f_2), G ) =0$. 
\end{definition}

Now we are in a position to present the Buchberger algorithm.
\begin{algorithm}{Buchberger}\\
For a given finite set $F \subset \kx$ determine the Groebner basis $G \subset \kx$
with $I(F)=I(G)$.\\[0pt]
\begin{center}
\colorbox{lgray}{\parbox{0.4\textwidth}
	{
\begin{align*}
&G:=F\\
&B:= \{ \{g_1, g_2\} | g_1, g_2 \in G \text{ with } g_1 \neq g_2 \}\\
&\text{\bf while } B \neq \emptyset \;\;\text{\bf do}\\
&	\di \text{choose }\{ g_1, g_2 \} \text{ from } B\\
&	\di B:=B \backslash \{\{g_1,g_2\}\}\\
&	\di h:= \spol(g_1, g_2)\\
&	\di h':= \normf(h,G)\\
&	\di {\bf if} \;\;h' \neq 0 \;\;\text{\bf then}\\ 
&	\di \di B:=B \cup \{\{g,h'\}|g \in G\}\\
&	\di \di G:=G \cup \{h'\}\\	
&	\text{\bf return } G
\end{align*}
	}}
\end{center}
\end{algorithm}

Note, that $G$ generates the same ideal as $F$, especially,
both sets have exactly the same simultaneous zeros.
It can be proven, that the Buchberger algorithm 
terminates as well as that the reduced Groebner basis 
is unique~\cite{Weispfenning}.
If we apply the Buchberger algorithm to 
the set~(\ref{eq-poly}) with subsequent reduction
we end up with the reduced Groebner basis (with
underlying lexicographical ordering)
\begin{align*}
g_1 &=y + x^2-3/2x-3,\\
g_2 &=x^3-5/2x^2-5/2x.
\end{align*}
The system of equations $g_1=g_2=0$ is equivalent to $f_1=f_2=f_3=0$, but the
former allows to directly calculate the solutions:
Since $g_2=0$ is univariate it can be solved immediately
and subsequently $g_1=0$ for each partial solution insertion.

Despite the correctness of the Buchberger algorithm tractability
of practical examples requires to improve this algorithm. 
In particular, the number of iterations in the algorithm
drastically grows with an increasing number 
of polynomials and with higher degrees of the 
polynomials.
In this respect much progress has been made with the improvement
of this original Buchberger algorithm 
from 1965~(see \cite{Weispfenning, Bose, Faugere-F4}).


\end{document}